\documentclass[12pt]{iopart}
\usepackage{graphics}

\def\etal{{\sl et al.\/}}
\def\etc{{\sl etc.\/}}
\def\ie{{\sl i.e.\/}}

\def\jhep{{\it J.\ H.\ E.\ P.\/},}
\def\jphys{{\it J.\ Phys.\/},}
\def\pr{{\it Phys.\ Rev.\/},}
\def\prl{{\it Phys.\ Rev.\ Lett\/},}
\def\pl{{\it Phys.\ Lett.\/},}

\begin{document}

\title{Phases and properties of quark matter}
\author{Sourendu Gupta (TIFR)\\Quark Matter 2008, Jaipur}
\address{Dept.\ of Theoretical Physics, Tata Institute of Fundamental
Research, Homi Bhabha Road, Mumbai 400005, India}
\ead{sgupta@theory.tifr.res.in}

\begin{abstract}
I review recent developments in finite temperature lattice QCD which are
useful for the study of heavy-ion collisions. I pay particular attention
to studies of the equation of state and the light they throw on conformal
symmetry and the large $N_c$ limit, and to the structure of the phase
diagram for $N_f=2+1$.
\end{abstract}

\maketitle

\section{Introduction}

Lattice gauge theory at finite temperature and density has been an
extremely active field in the last year. This makes it hard to review
it in its entirety within the scope of this article. I have therefore
chosen to review two of the major points of contact with heavy-ion
physics, namely the equation of state and the phase diagram, in the
later sections. In this section I provide pointers to the literature on
the many developments I will not discuss further.

\begin{itemize}
\item The finite-temperature phase transition is now definitely established
  to be a cross over: this is verified by the Budapest-Wuppertal (BW) group 
  \cite{bw,bw2}, and confirmed by the BBRC collaboration \cite{bbrc,bbrc2}.
  Further confirmation has come from the Hot-QCD collaboration.
  The cross over temperature is  temporarily in dispute. The 
  Budapest-Wuppertal group \cite{bw2} finds a cross-over temperature
  substantially smaller than that obtained by BBRC \cite{bbrc2} using the
  same thermometer. An old global analysis in 2001 \cite{oldglobal} gave
  $T_c\simeq175$ MeV with 20 MeV uncertainty from scale setting. BBCR and
  Hot-QCD prefer the upper end, BW prefer the lower end. This disagreement,
  while serious, therefore does not impact the experimental search.
\item Correlation function measurements have led to new developments.
  Deconfinement occurs at the chiral cross over point: this has been
  shown using linkages between quantum numbers; for example, the linkage
  between baryon number and strangeness becomes exactly that expected from
  quarks at $T_c$ \cite{linkage}.
  There is steady and slow advance in the measurement of transport
  coefficients on the lattice \cite{transport}.
  Renormalized Polyakov loop measurements in various representations
  give strong evidence for Casimir scaling to all orders \cite{cas}. The
  octet loop, in particular, does not see the phase transition.
  The non-melting of $J/\psi$ and the melting of $\chi_c$ soon above
  $T_c$ is now verified in many different computations \cite{nonmelt}.
\item There has been significant advance in defining chiral fermions at
  finite chemical potential \cite{chiral}. There are advances in the
  understanding of isospin chemical potential and imaginary chemical
  potential \cite{isoimagmu}.
  There are new results for thermodynamics using Wilson quarks
  \cite{wilsonth}.
  Localization of staggered Dirac eigenvectors is seen to set in abruptly
  at $T_c$, although this could be a finite volume artifact \cite{local}.
\end{itemize}

\section{The equation of state}

\begin{figure}
\begin{center}
\scalebox{0.5}{\includegraphics{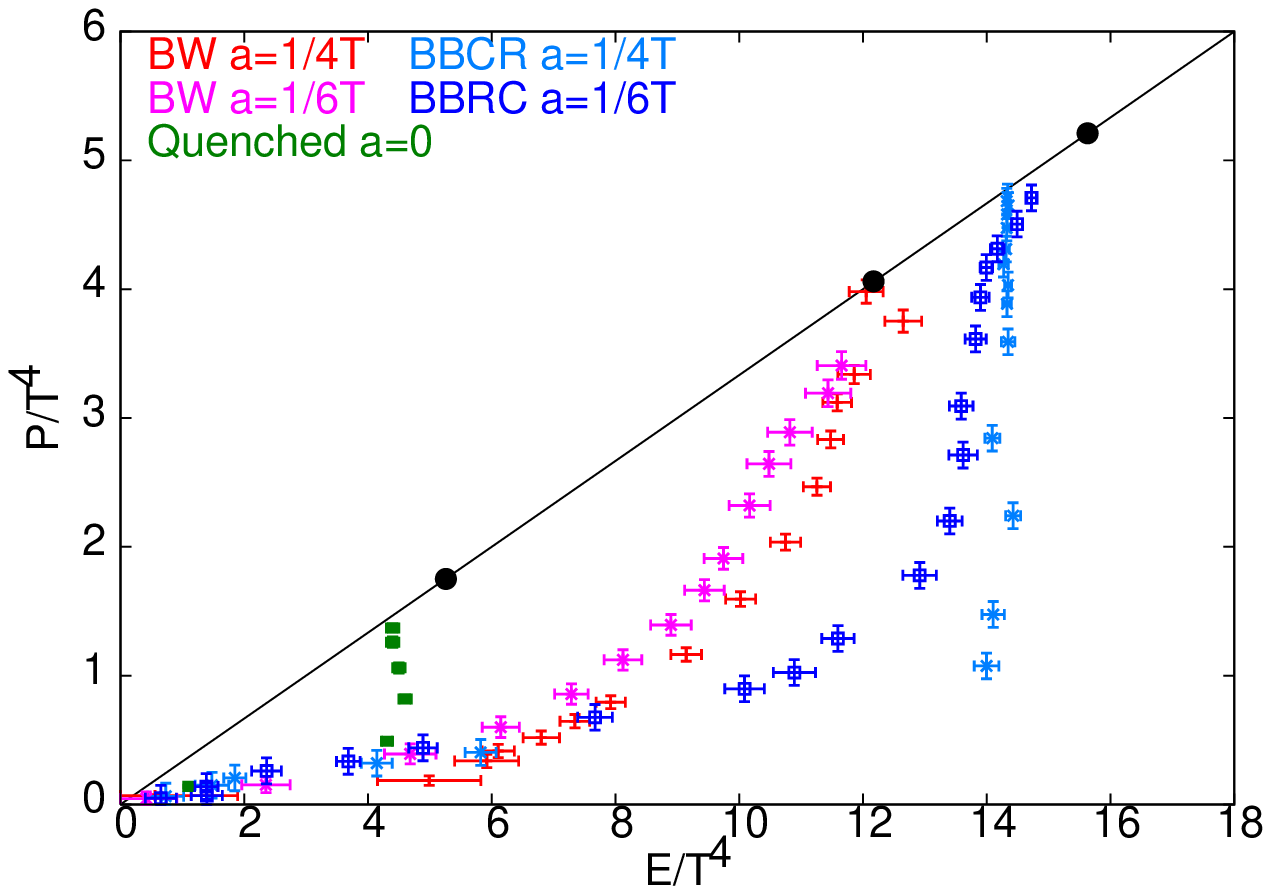}}
\scalebox{0.5}{\includegraphics{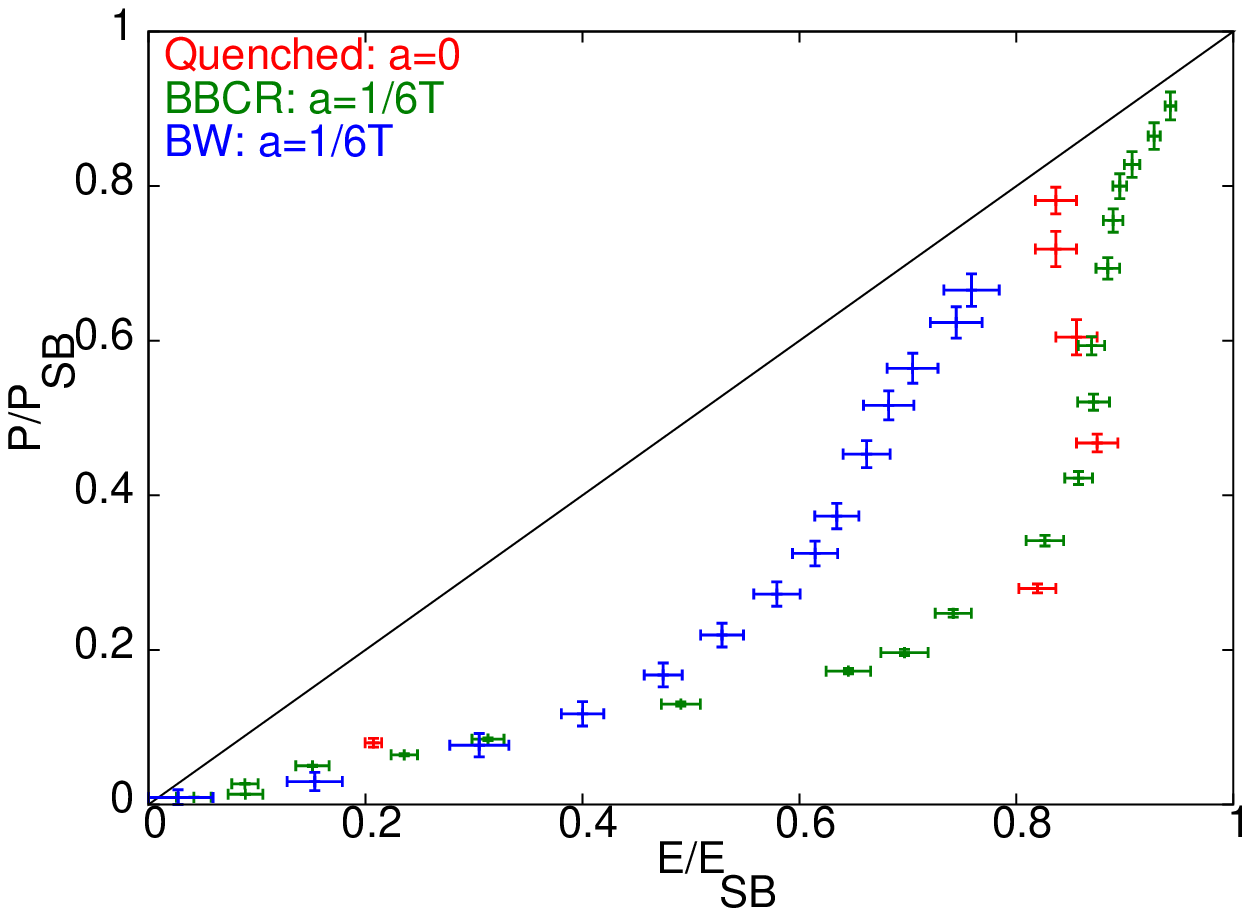}}
\end{center}
\caption{Quenched data from \cite{gavaithermo}, $N_f=2+1$ data from
   \cite{bw2,bbrc2}. On the left, is a plot of $P/T^4$ against $E/T^4$. On
   the right is a plot of $P/P_{SB}$ and $E/E_{SB}$. The quenched data is
   scaled by the $N_f=0$ continuum SB values, $N_f=2+1$ data is scaled by
   the $N_f=3$ continuum SB values.}
\label{fg.eos}
\end{figure}

The equation of state is rather well-known for QCD without quarks. Recently
there have been three separate efforts to extract it for $N_f=2+1$,
\ie, two light degenerate flavours (corresponding to $u$ and $d$ quarks)
and one heavier flavour. The BW collaboration uses staggered quarks and
sets the renormalized quark masses to their physical values by tuning
to realistic values of $m_\pi$ and $m_K$. The BBRC collaboration uses
P4 quarks with light quark masses heavier than physical but a physical
strange quark masses by tuning to a heavier pion ($m_\pi\simeq220$ MeV)
and a realistic $m_\phi$. The HotQCD collaboration is investigating the
equation of state with both P4 and Asqtad quarks at renormalized quark
masses equal to that used by the BBRC collaboration.

The equation of state of state in the form $P(E)$ is an important input
into hydrodynamics, and therefore very important for predictions of
various signals in heavy-ion collisions.  In conformal field theories,
including free field theory, one necessarily has $P=E/3$, so that
$c_s^2=1/3$.  This limit has excited interest recently, since many
toy models for QCD can be solved using AdS/CFT techniques which demand
conformal symmetry along with the large $N_c$ approximation.

The lattice data is displayed in the two composite plots in Figure
\ref{fg.eos}.  The diagonal line denotes conformal equations of state:
points corresponding to the ideal gases for $N_f=0$, 2 and 3 are marked on
this line.  Note that the data on the quenched theory is extrapolated to
the continuum limit, whereas those for $N_f=2+1$ are for a finite lattice
cutoff $a=1/4T$ and $a=1/6T$. There is apparent disagreement between the
computations with staggered and P4 quarks: they are not expected to agree
at finite lattice spacing, only in the continuum limit must they give
the same result. In this connection note that the results of the HotQCD
collaboration at $a=1/8T$ smoothens the behaviour of the P4 quarks,
reducing the maximum excursion from the conformal line by about 20\%,
while leaving data at $3T_c/3$ and higher almost unchanged. In summary,
there is clear evidence for strong violation of conformal symmetry in
the region of the crossover from hadron to quark phases.

Importantly, the deviations are not restricted to a narrow region around
$T_c$, but extend to $T=2T_c$ or slightly higher. This will have two
consequences in applications to heavy-ion collisions---
\begin{enumerate}
\item Hydrodynamics of non-ideal fluids with conformal equations of state
($P=E/3$) allow only for shear viscosity. If conformal symmetry is broken,
the fluid can have bulk viscosity as well. Closely related to this is the
fact that $c_s^2<1/3$.
\item Unless initial temperatures rise well above $2T_c$, heavy-ion collisions
may never be, {\sl a priori\/}, approximated as systems with vanishing bulk
viscosity.
\end{enumerate}

\begin{figure}
\begin{center}
\scalebox{0.6}{\includegraphics{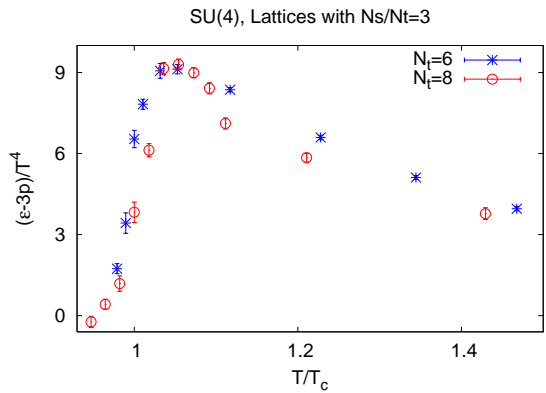}}
\end{center}
\caption{Strong deviations from conformal symmetry are observed for
 $N_c=4$ and 6 \cite{sunc}.}
\label{fg.sunc}
\end{figure}

A recent study of SU($N_c$) pure gauge theories shows that this
behaviour persists for $N_c>3$ \cite{sunc}. The breaking of
conformal symmetry is often exhibited as a plot of $E-3P$ against $T$
(see Figure \ref{fg.sunc}). For SU(3) gauge theory with or without quarks,
peaks in this quantity are seen close to $T_c$. For pure gauge theories,
there is a first order phase transition, and the maximum value of $E-3P$
is at least as large as the latent heat density, and hence expected
to scale at $N_c^2$. Since the Stephan-Boltzmann values of $E$ and $P$
also scale as $N_c$, the breaking of conformal symmetry for SU($N_c$)
pure gauge theories is also expected to persist to large $N_c$.

In heavy-ion collisions a very important use of the equation of state is the
estimate of typical energy densities in the plasma phase and in the hadronic
phase of QCD. For reference I quote some estimates for the energy density
from current lattice computations just below the lowest estimate of the
crossover temperature, and just above the highest estimate.
We can take as fiducial values $E_1=E(140{\rm\ MeV})$ which is definitely in
the hadron phase and $E_2=E(210{\rm\ MeV})$ which is definitely in the plasma
phase. The current estimates are
\begin{equation}
   E_1 = \cases{
     70\pm12\;\; {\rm MeV}/{\rm fm}^3 \quad&\quad(BBRC, $1/a=$840 MeV),\\
     21\pm24\;\; {\rm MeV}/{\rm fm}^3 \quad&\quad(BW, $1/a=$840 MeV).}
\label{referelo}
\end{equation}
In the high temperature phase we have,
\begin{equation}
   E_2 = \cases{
     3.11\pm0.07\;\; {\rm GeV}/{\rm fm}^3 \quad&\quad(BBRC, $1/a=$1260 MeV),\\
     2.28\pm0.08\;\; {\rm GeV}/{\rm fm}^3 \quad&\quad(BW, $1/a=$1260 MeV).}
\label{referehi}
\end{equation}
While noting the statistically significant differences between the results of
the two computations, note also that the renormalized quark masses differ,
and that both estimates are made at finite lattice spacing with quark
formulations which differ at finite cutoff.
For comparison, note that in the continuum limit of quenched QCD one finds
$E(T_c=285{\rm\ MeV})=3.4\pm0.5$ Gev/fm${}^3$ \cite{gavaithermo}.

The success of a hydrodynamic description of heavy-ion collisions would
eventually be gauged from its ability to reliably extract $P(E)$ (and
other material properties such as transport coefficients) from data,
thus permitting experimental tests of lattice predictions. Any claims
of disagreement between such extractions and lattice predictions are
actually claims of the failure of QCD, and therefore should be treated
with the strong, fair and rational skepticism that any claim of a failure
of a well-established theory should receive. At this point of time 
it is easier to control lattice studies: one merely has to spend computer
time in decreasing the lattice spacing, a procedure that is already under
way. Control of the hydrodynamic description is more problematic, with
initial conditions, the hydrodynamic equations, and the hadronization
prescription each still in need of independent validation.

\section{The global phase diagram}

\begin{figure}
\begin{center}
\scalebox{0.5}{\includegraphics{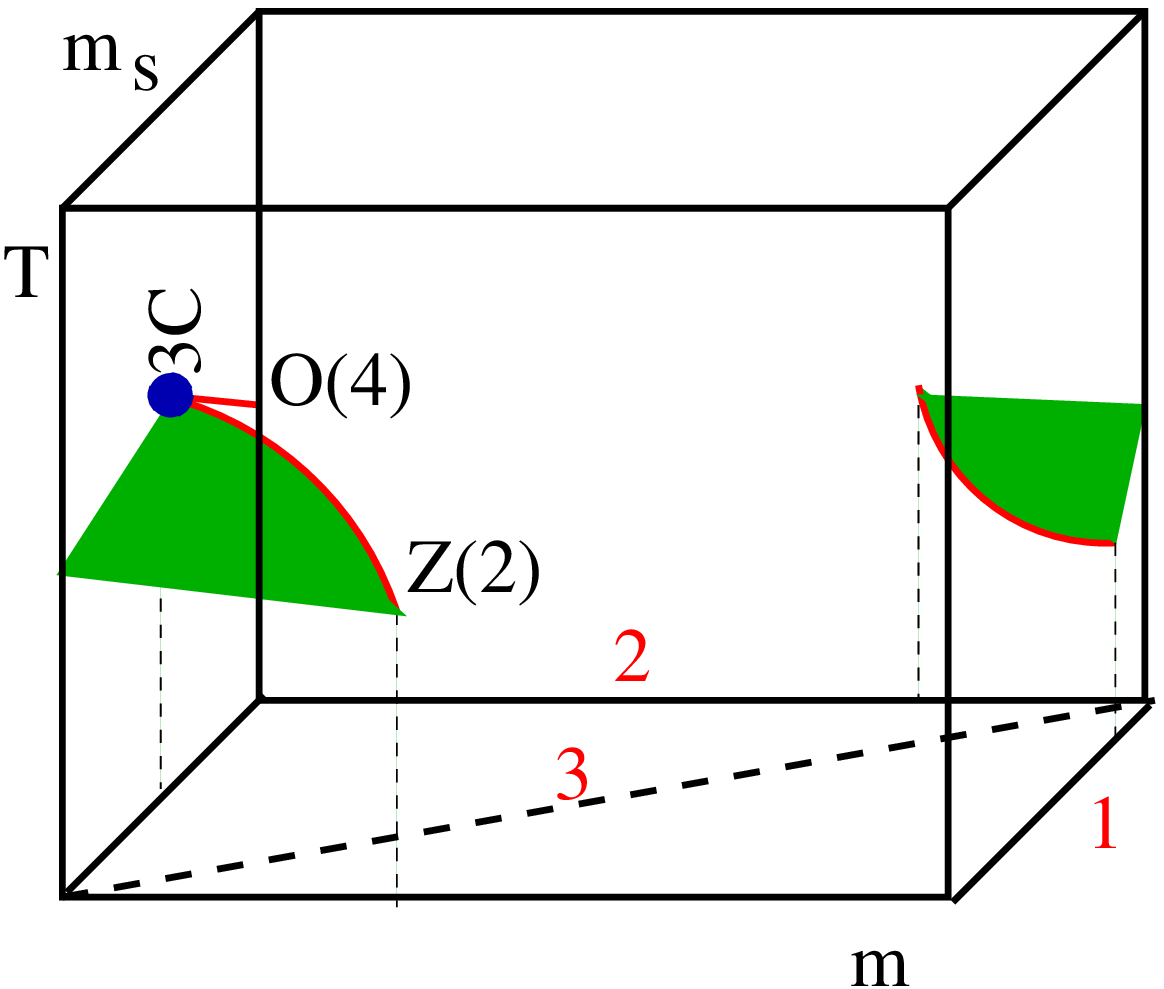}}
\scalebox{0.54}{\includegraphics{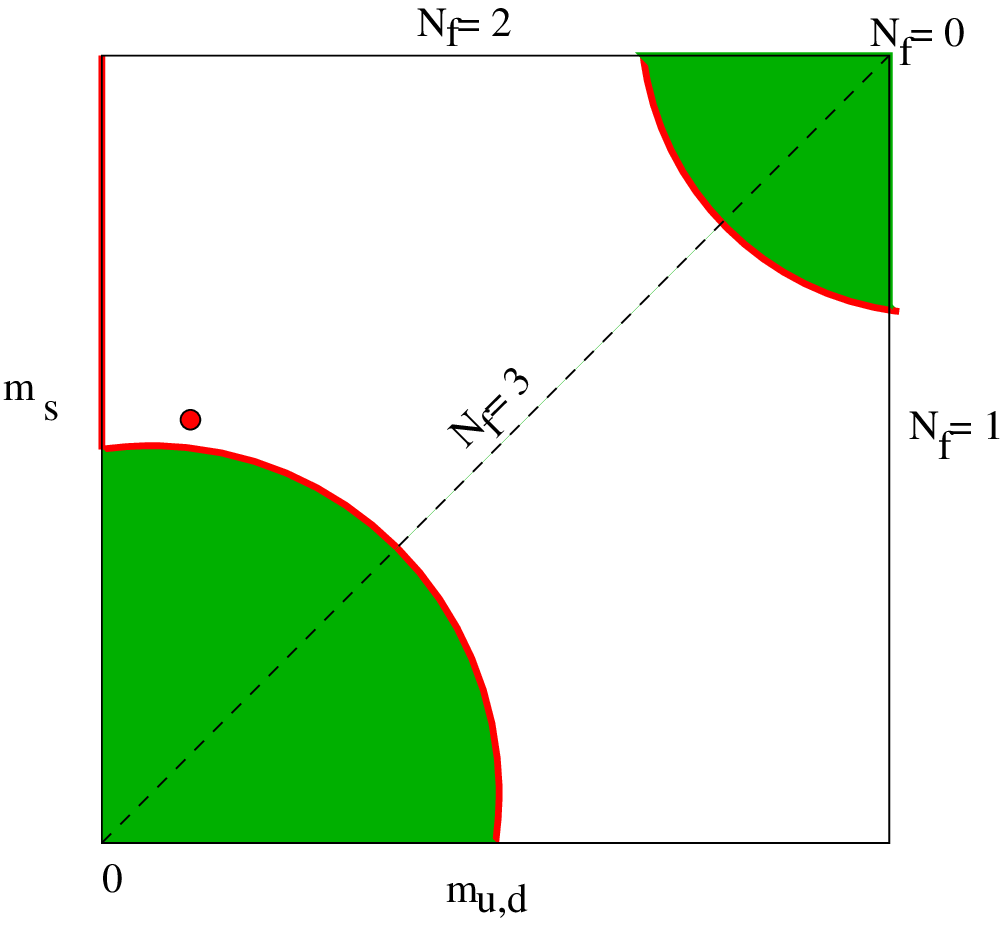}}
\end{center}
\caption{The phase diagram on the left is put together from computations by
  \cite{columbia} and arguments from \cite{piwi}.
  Projecting the phase diagram down to the $T=0$ plane, one obtains the
  flag diagram, which is the figure on the right.}
\label{fg.flag}
\end{figure}

Phase diagrams are labeled by the thermodynamic intensive
coordinates. For QCD these are $T$, $N_f$ quark masses and $N_f$
chemical potentials. Of these, experiments can tune (at best) $1+N_f$
of these, since the quark masses are fixed conditions that we are faced
with in reality. In practice, heavy-ion collisions have only a single
control parameter--- $\sqrt S$. This is not sufficient to examine the
4 dimensional phase diagram of QCD: only enough to explore a single
line through the phase diagram.  By varying the ions one can smear
this line a bit, but this still leaves scope for much thought and
experimental ingenuity in exploring larger parts of the phase diagram
in the laboratory.

Each point in phase diagram is, almost always, a single pure phase.
Exceptions are where two or more phases coexist; these are also called
first order transitions. A continuity argument for lines (surfaces)
of first order transition is called the Gibbs' phase rule. It follows
from the structure of the solutions of the equation $g_A(T,\mu_i,m_i)
= g_B(T,\mu_i,m_i)$, where $g_{A,B}$ are the free energy density in
the two phases $A$ and $B$.  The Gibbs' phase rule implies that in D
dimensional phase diagram one has D-2 dimensional critical surfaces,
D-3 dimensional tricritical surfaces, D-4 dimensional tetracritical
surfaces \etc.  This argument, and equivalent forms of it, strongly
constrains the topology of phase diagrams \cite{sgphsd}.

When discussing high dimensional phase diagrams, one often discusses sections,
\ie, parts of the phase diagram with some of the intensive variables set equal
to fixed values. These are also phase diagrams, in the sense that each point
corresponds to an unique thermodynamic phase. However, in QCD it has become the
practice to use projections, which we call flag diagrams. A construction of a
flag diagram is shown in Figure \ref{fg.flag}. Each point in a flag diagram
shows the kind of phase transition that one obtains by varying the quantity
which has been projected away.

\begin{figure}
\begin{center}
\scalebox{0.5}{\includegraphics{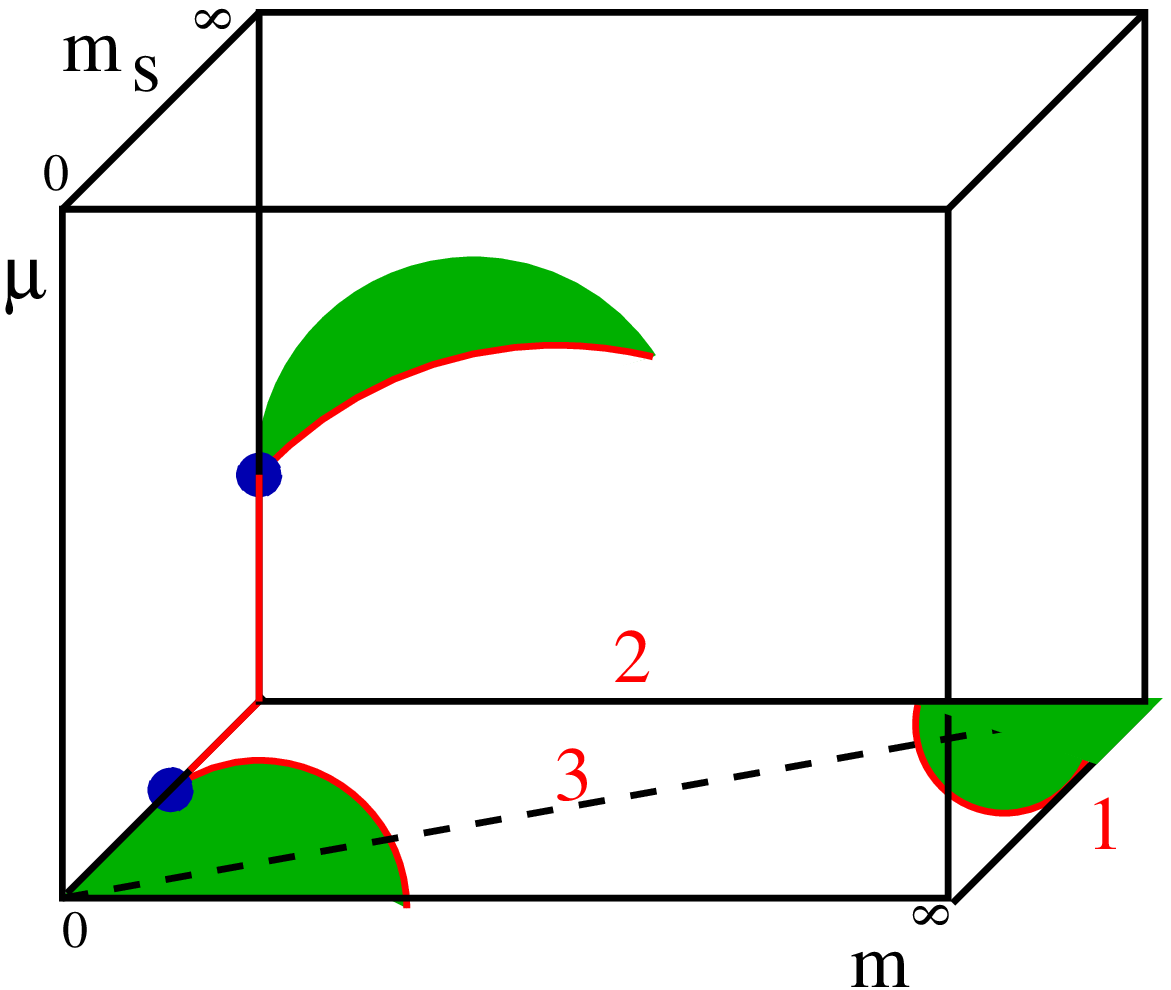}}
\scalebox{0.5}{\includegraphics{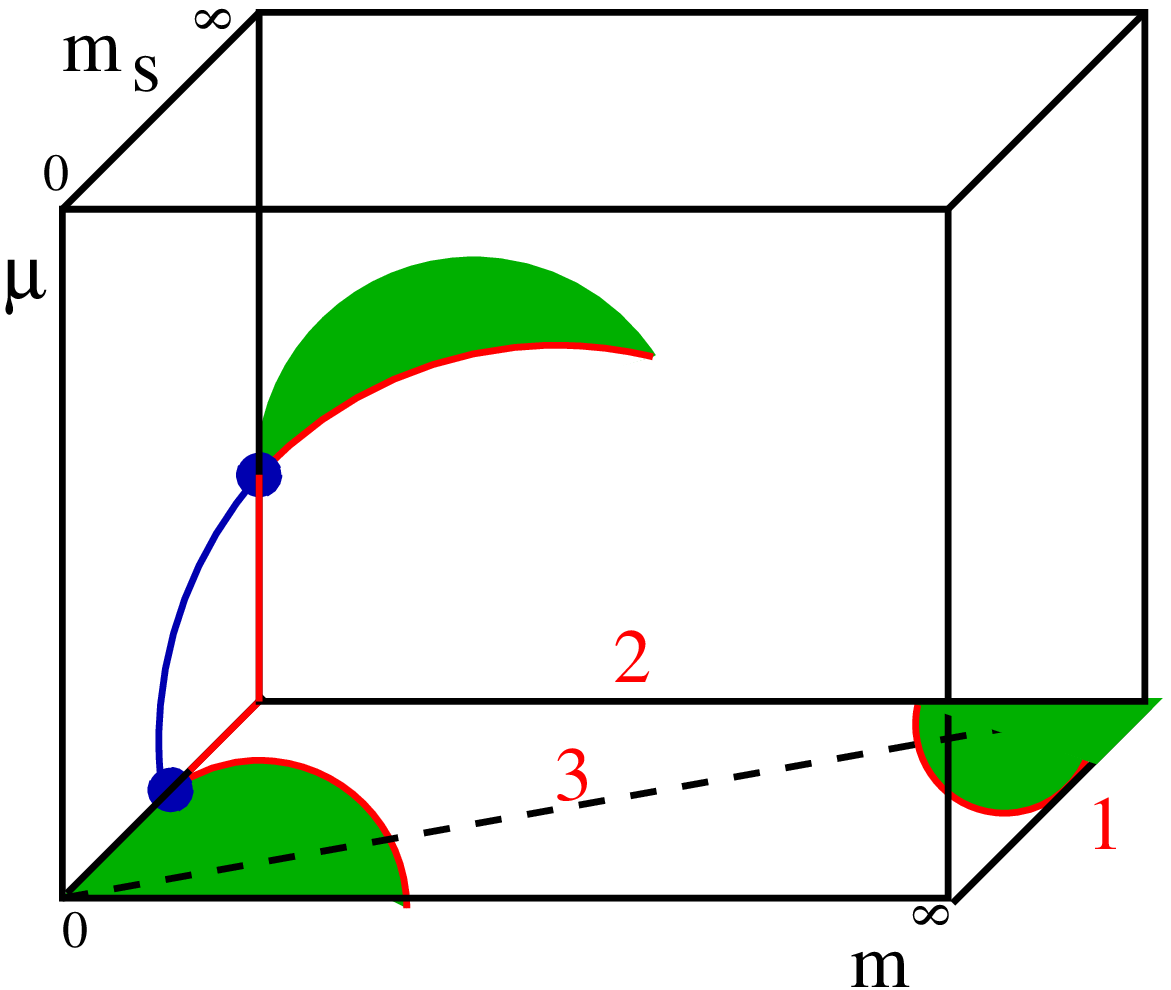}}
\scalebox{0.5}{\includegraphics{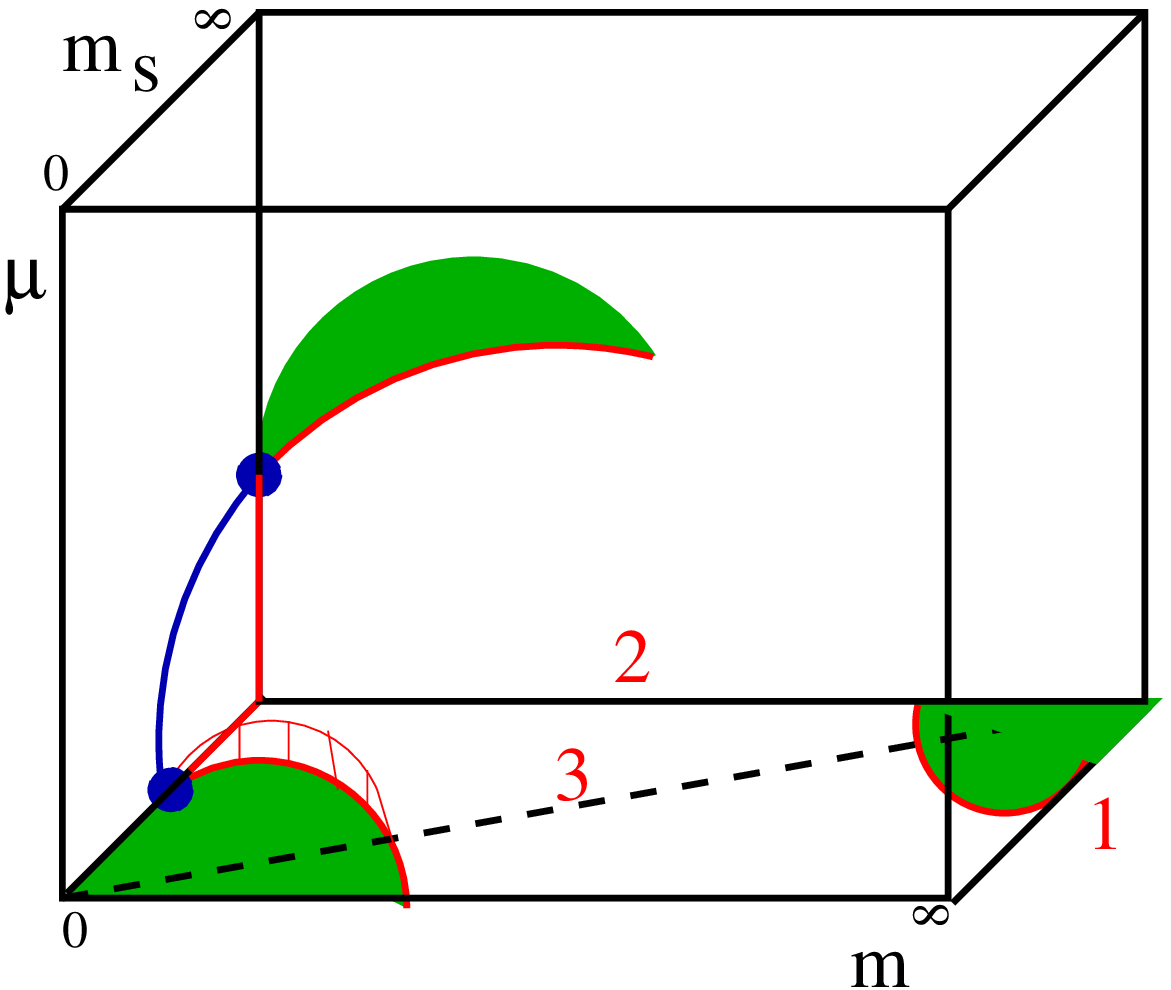}}
\scalebox{0.5}{\includegraphics{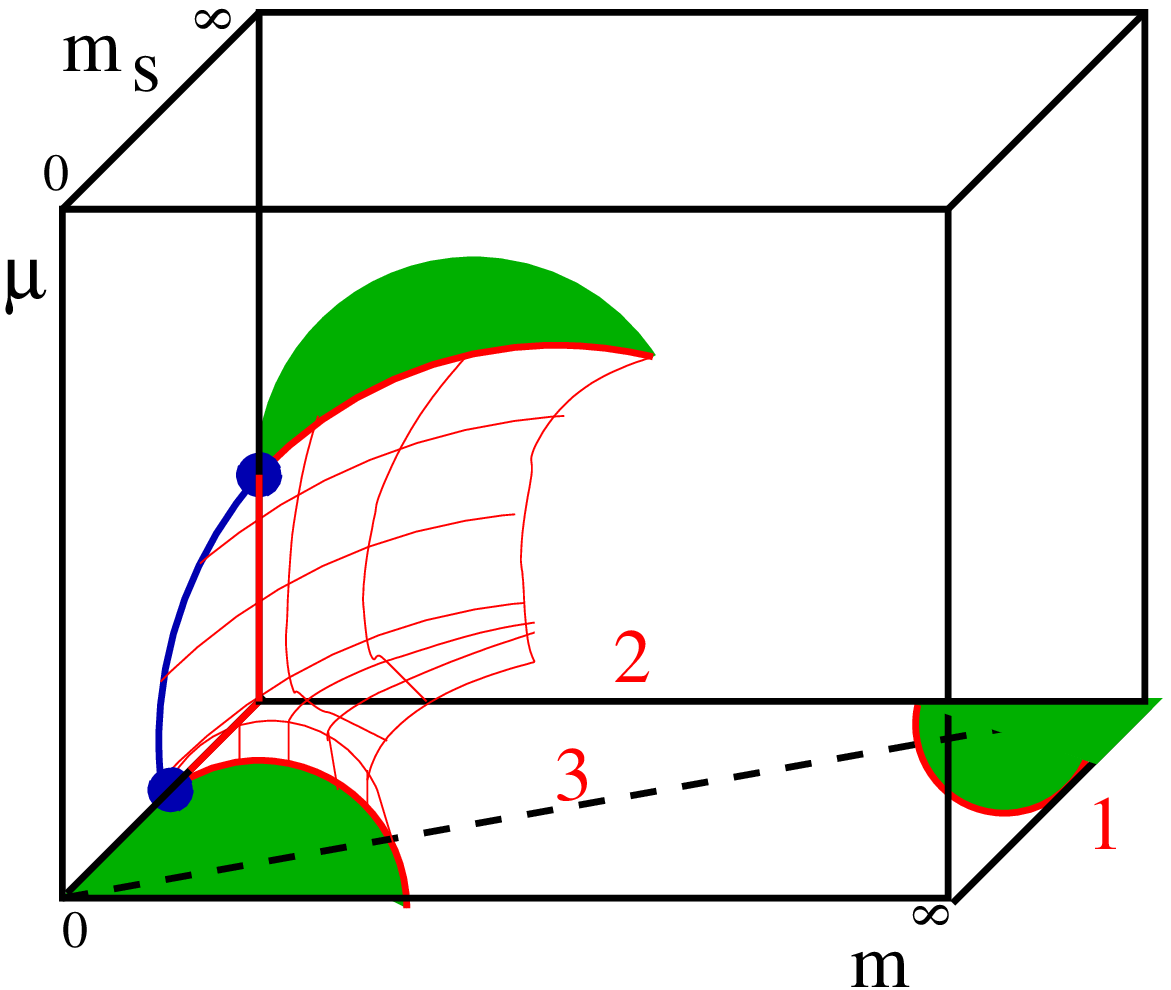}}
\end{center}
\caption{The extended flag diagram of QCD. The planes $\mu=0$ and $m_s=\infty$
  of this flag diagram are reasonably well explored, and have the structure
  shown in the first panel. There are two tricritical points, at each of which
  an Ising critical line joins an O(4) critical line. Since the two O(4)
  critical lines bound an O(4) critical surface, the two tri-critical points
  are joined together by a tri-critical line. This line has to lie on the
  $m=0$ plane, since the the O(4) surface lies in that plane. In \cite{defp}
  it was found that a part of the Ising critical surface near the $N_f=2+1$
  region bends ``backwards''. Since a critical surface does not have ``surface
  tension'', it can twist and bend. To look at the possible phase diagrams,
  examine the shape of the tricritical line.}
\label{fg.fullflag}
\end{figure}

The QCD flag diagram can be extended to include finite baryon chemical
potential $\mu$. Note that we are discussing a 3-d flag diagram obtained
from the 4-d section of the full phase diagram with $m_u=m_d=m$, and the
two chemical potentials, $\mu_s=\mu_I=0$. In this flag diagram the planes
of $m_s=\infty$ and $\mu=0$ have been explored, and the topology of
the phase diagrams are reasonably well understood. In each of these planes
there is a tricritical point at which an Ising critical line joins on to
an O(4) critical line. One of these tricritical points is shown in Figure
\ref{fg.flag}. Now, the two O(4) critical lines are just the boundary of a
single O(4) critical surface. Since these two end in a tricritical point,
there must be a tricritical line bounding the O(4) critical surface, of which
these two points are the ends (see Figure \ref{fg.fullflag}).

Along this tricritical line one must glue together an O(4) critical
surface and one or more Ising critical surfaces. Now, \cite{defp} found
that a part of the Ising critical surface near the $N_f=2+1$ region bends
in the direction of smaller $m_s$. With this evidence they claimed that
the tricritical line must occur in two pieces, and the two known Ising
critical lines must belong to two different Ising critical surfaces. This
argument, while compelling, is not water-tight. The reason is that there
is no physical principle that constrains the curvature of critical (and
tricritical) surfaces: they are allowed to bend, twist and wander. Hence
the ``wrong'' curvature seen in \cite{defp} can be accommodated into the
flag diagram shown in Figure \ref{fg.fullflag}. One must enumerate all
topologies that the tricritical line may have.

\begin{figure}
\begin{center}
\scalebox{0.33}{\includegraphics{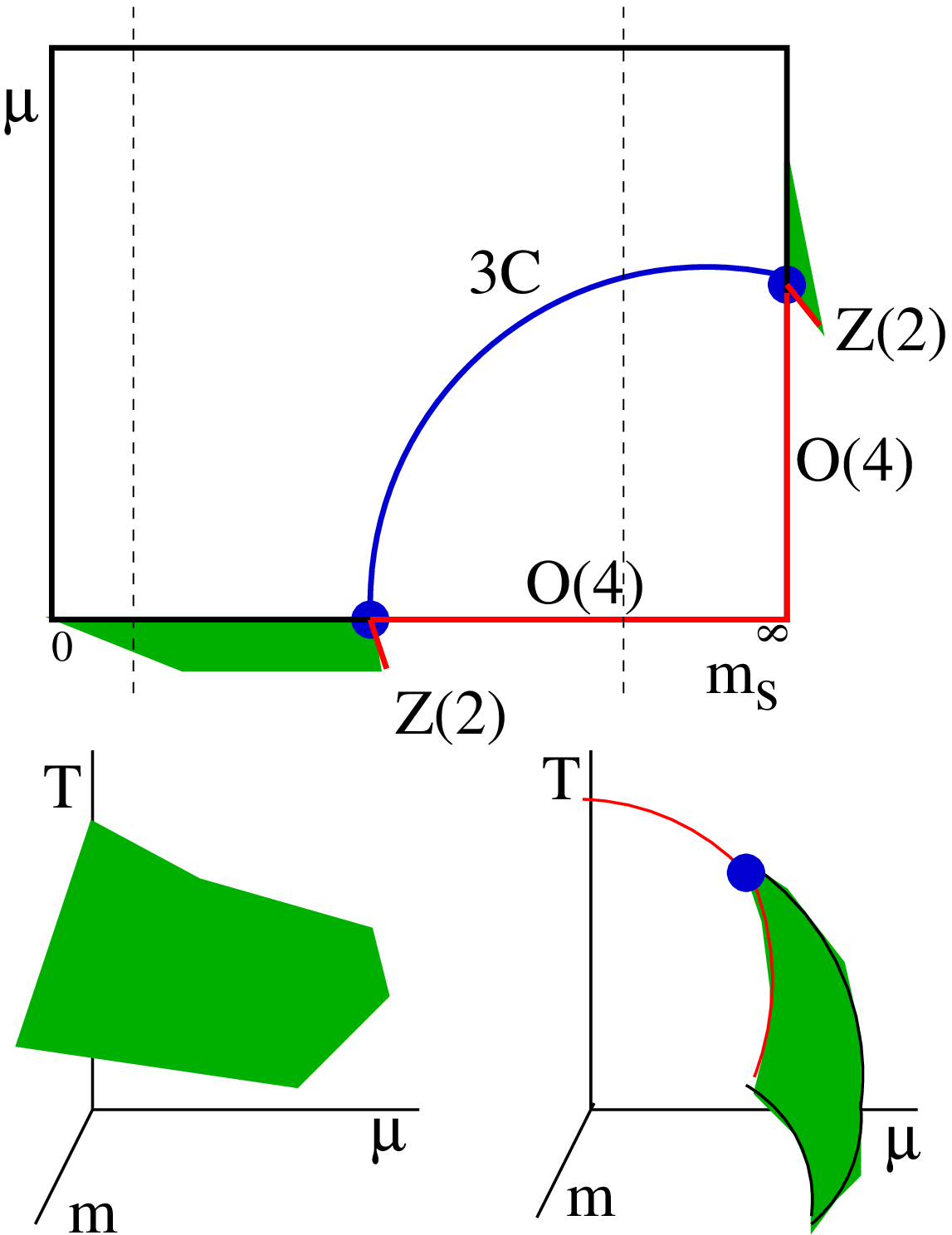}}\hfill
\scalebox{0.33}{\includegraphics{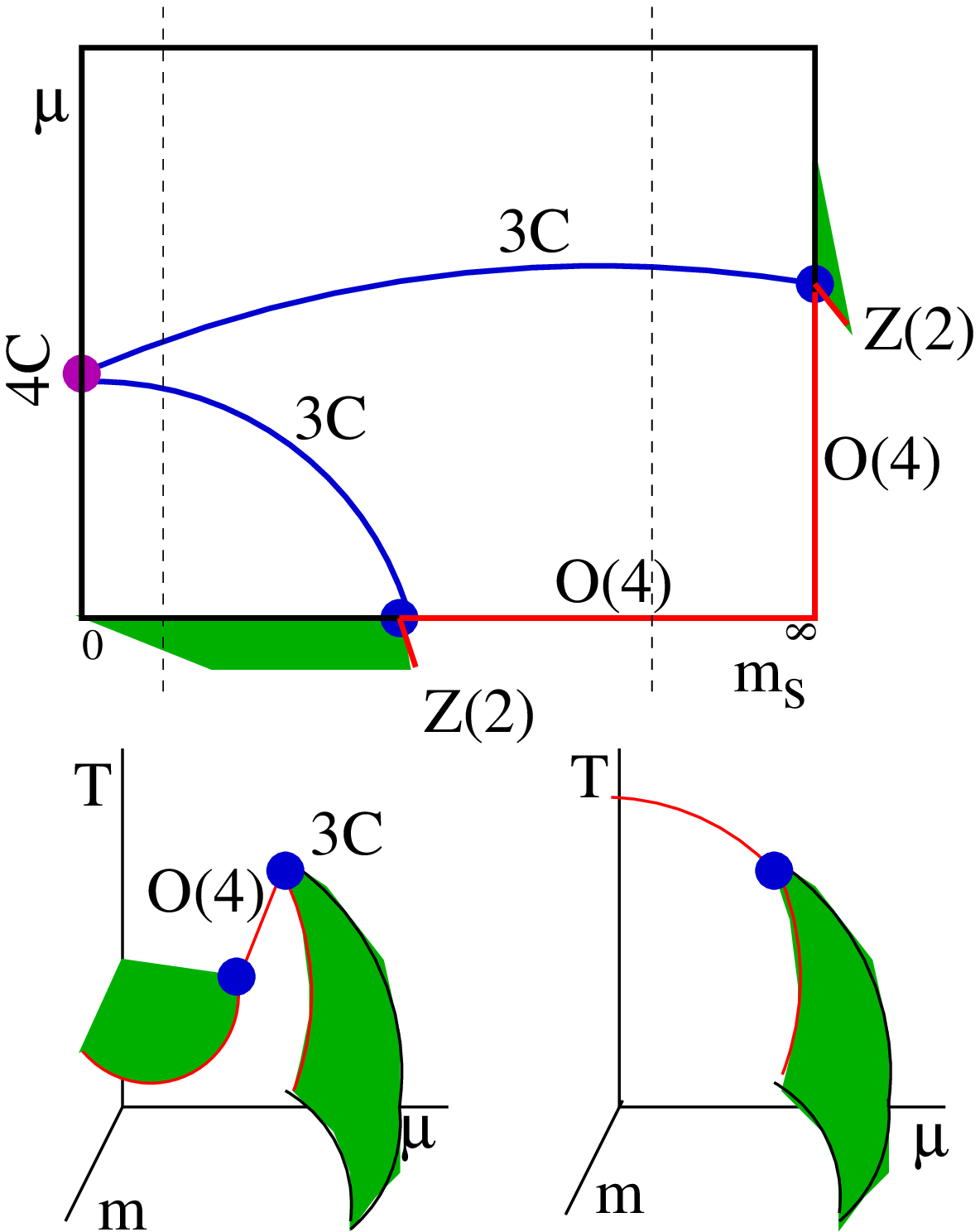}}\hfill
\scalebox{0.33}{\includegraphics{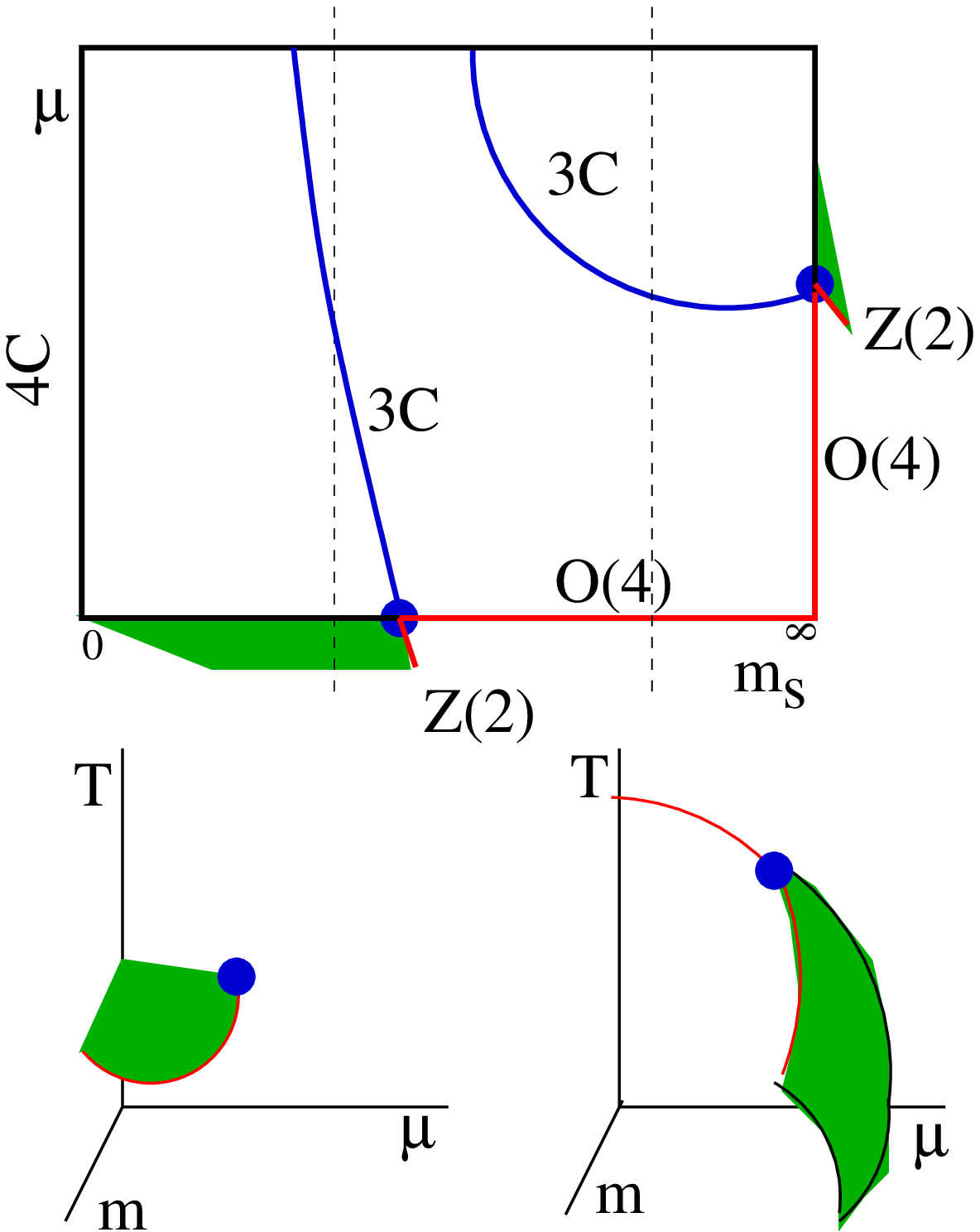}}
\end{center}
\caption{Three topologies for the tricritical line are shown. Below each
  is shown the phase diagram corresponding to two different generic quark
  masses. Note that the second case above is ruled out by universality
  arguments.}
\label{fg.tricritical}
\end{figure}

Apart from the simple topology shown in the last panel of Figure
\ref{fg.fullflag}, there are several more possibilities. In a 4-d phase
diagram one non only has the possibility of a tricritical line, but also
a tetracritical point where two such tricritical lines meet. If indeed the
two Ising critical surfaces are physically distinct, then the possibility
of a tetracritical point must be considered. Usually, one would discover
a higher-order critical behaviour at a point of enhanced symmetry. Such
points are along the 3 flavour line $m_s=m_u=m_d$. However, this is ruled
out, since universality arguments rule out critical behaviour for SU(3)
flavour.  The remaining possibility is to draw the tetracritical point
out to infinity.  These three distinct possibilities for the topology
of the tricritical line] are shown in Figure \ref{fg.tricritical}, along
with the three dimensional physical phase diagrams that they would give
rise to at generic quark masses. Note that the second case is ruled out
by universality.  Which of these cases is true for QCD is something that
has to be decided by further lattice simulations.

\section{Summary}

Significant progress has been made in lattice computations in the last
year.  The new state of the art is $m_\pi\simeq140$--220 MeV. For the
first time all collaborations are using $m_\pi<m_\rho/2$. Conformal
symmetry is strongly broken for $T<2T_c$, as shown directly by the
equation of state. There is direct evidence from the lattice that
conformal symmetry breaking persists for larger number of colours,
and hence, through power counting arguments, to all $N_c$. One major
consequence of this observation is that hydrodynamic studies should
include bulk viscous terms. I have argued that a QCD critical point
could exist at small chemical potentials even in the $N_f=2+1$ theory,
and that it is still possible that $N_f=2$ computations are a good
guide to its location. Further lattice computations are called for.

\ack
It is a pleasure to thank my collaborators, Saumen Datta and Rajiv
Gavai, and also Philip de Forcrand, Rajan Gupta, Frithjof Karsch and
Bengt Petersson for discussions. I would like to thank especially the
BW and BBRC collaborations for the use of their results.


\begin{thebibliography}{99}
\bibitem{bw} Aoki Y \etal, {\it Nature\/}, {\bf 443} (2006) 675.
\bibitem{bw2} Aoki Y \etal, \jhep {\bf 0601} (2006) 089.
\bibitem{bbrc} Cheng M \etal, \pr {\bf D 74} (2006) 054507.
\bibitem{bbrc2} Cheng M \etal, \pr {\bf D 77} (2008) 014511.
\bibitem{oldglobal} Gupta S, \pr {\bf D 64} (2001) 034507.
\bibitem{linkage} Gavai R V and Gupta S, \pr {\bf D 73} (2006) 014004.
\bibitem{transport} Aarts G \etal, \prl {\bf 99} (2007) 022002;
  Meyer H B, arXiv:0710.3717.
\bibitem{cas} Gupta S, Hubner K and Kaczmarek O, \pr {\bf D 77} (2008) 034503.
\bibitem{nonmelt} D\"oring M \etal, \pr {\bf D 75} (2007) 054504;
  Aarts G \etal, \pr {\bf D 76} (2007) 094513;
  Umeda T arXiv:0710.0204.
\bibitem{chiral} Gattringer C \etal, \pr {\bf D 76} (2007) 054502;
  Banerjee D, Gavai R V and Sharma S, arXiv:0803.3295.
\bibitem{isoimagmu} Splittorff K and Svetitsky B, \pr {\bf D 75} (2007) 114504;
  Conradi S and d'Elia M, \pr {\bf D 76} (2007) 074501;
  Kogut J B and Sinclair D K, arXiv:0712.2625;
  Cea P \etal, \pr {\bf D 77} (2008) 051501.
\bibitem{wilsonth} Maezawa Y \etal, \jphys, G 34 (2007) S651;
  Chen H S and Luo X Q, hep-lat/0702025;
  Creutz M, \pr {\bf D 76} (2007) 054501.
\bibitem{local} Gavai R V , Gupta S and Lacaze R, arXiv:0803.0182.
\bibitem{gavaithermo} Gavai R V, Gupta S and Mukherjee S, hep-lat/0506015.
\bibitem{hotqcd} Detar C \etal, arXiv:0710.1655.
\bibitem{sunc} Datta S and Gupta S, in progress.
\bibitem{defp} de Forcrand P and Philipsen O, \jhep, 0701 (2007) 077.
\bibitem{columbia} Brown F R \etal, \prl {\bf 65} (1990) 2491;
  Karsch F, Laermann E and Schmidt C, \pl {\bf B 520} (2001) 41.
\bibitem{piwi} Pisarski R D and Wilczek F, \pr {\bf D 29} (1984) 338.
\bibitem{sgphsd} Gupta S, arXiv:0712.0434.
\end{thebibliography}
\end{document}